# Experimental model of topological defects in Minkowski spacetime based on disordered ferrofluid: magnetic monopoles, cosmic strings and the spacetime cloak

Igor I. Smolyaninov [1], Vera N. Smolyaninova [2], Alexei I. Smolyaninov [3],

[1] *Department of Electrical and Computer Engineering, University of Maryland, College Park, MD 20742, USA*

[2] *Department of Physics Astronomy and Geosciences, Towson University,*

*8000 York Rd., Towson, MD 21252, USA*

[3] *Department of Electrical and Computer Engineering, University of California, San Diego, La Jolla, CA 92093, USA*

**Cobalt nanoparticle-based ferrofluid in the presence of external magnetic field forms a self-assembled hyperbolic metamaterial. Wave equation describing propagation of extraordinary light inside the ferrofluid exhibits 2+1 dimensional Lorentz symmetry. The role of time in the corresponding effective 3D Minkowski spacetime is played by the spatial coordinate directed along the periodic nanoparticle chains aligned by the magnetic field. Here we present a microscopic study of point, linear, planar and volume defects of the nanoparticle chain structure and demonstrate that they may exhibit strong similarities with such Minkowski spacetime defects as magnetic monopoles, cosmic strings and the recently proposed spacetime cloaks. Experimental observations of such defects are described.**

Metamaterials are artificial materials built from conventional microscopic materials in order to engineer their properties in a desired way. Such artificial metamaterials offer novel ways for controlling electromagnetic, acoustic, elastic and thermal properties of matter. However, considerable difficulties still exist in fabrication of three-dimensional metamaterials. Therefore, they are typically confined to small sizes in two dimensions. Recent demonstration that cobalt nanoparticle-based ferrofluid subjected to external magnetic field exhibits hyperbolic metamaterial properties [1,2] opens up many new directions in transformation optics and metamaterial research, since such a metamaterial is created by 3D self-assembly, and its dimensions are not limited by nanofabrication issues. On the other hand, 3D self-assembly of macroscopic metamaterial samples unavoidably leads to formation of various defects in the bulk of the metamaterial, which may affect its optical properties. While very important from the practical standpoint, this issue is also very interesting for purely scientific reasons. It appears that the wave equation describing propagation of extraordinary light inside the hyperbolic metamaterial exhibits 2+1 dimensional Lorentz symmetry, while the role of time in the corresponding effective 3D Minkowski spacetime is played by the spatial coordinate directed along the optical axis of the metamaterial [3,4]. Therefore, point, linear and planar defects of the self-assembled nanoparticle chain structure may exhibit strong similarities with such Minkowski spacetime defects as cosmic strings [5] and magnetic monopoles [6]. On the other hand, volume defects of the aligned nanoparticle chain structure may correspond to the recently proposed spacetime cloak geometry [7].

Modern developments in gravitation research strongly indicate that classic general relativity is an effective macroscopic field theory, which needs to be replaced with a more fundamental theory based on yet unknown microscopic degrees of freedom. However, our ability to obtain experimental insights into the future fundamental theory is strongly limited by low energy scales available to terrestrial particle physics and astronomical observations. The emergent analogue spacetime program offers a promising way around this difficulty. Looking at such systems as superfluid helium and cold atomic Bose-Einstein condensates, physicists learn from Nature and discover how macroscopic field theories arise from known well-studied atomic degrees of freedom. Ferrofluid in external magnetic field provides us with yet another easily accessible experimental system, which lets us trace the emergence of effective Minkowski spacetime from well understood microscopic degrees of freedom. Here we will describe direct experimental observations of microscopic defects of the effective Minkowski spacetime and discuss their influence on the hyperbolic metamaterial properties of the ferrofluid. As pointed out recently by Mielczarek [8], the properties of self-assembled magnetic nanoparticle-based hyperbolic metamaterials exhibit strong similarities with the properties of some quantum gravity models, such as loop quantum cosmology. Moreover, the physical vacuum appears to exhibit hyperbolic metamaterial properties when subjected to a very strong magnetic field [9,10]. Therefore, our experimental observations have many important fundamental implications, which reach far beyond transformation optics and electromagnetic metamaterial theory.

Our experimental technique uses three dimensional self-assembly of cobalt nanoparticles in the presence of external magnetic field. Magnetic nanoparticles in a ferrofluid are known to form nanocolumns aligned along the magnetic field [11]. Moreover, depending on the magnitude of magnetic field, nanoparticle concentration and solvent used, phase separation into nanoparticle rich and nanoparticle poor phases may occur in the ferrofluid [1]. This phase separation occurs on a 0.1-1 micrometer scale. For our experiments we have chosen cobalt magnetic fluid 27-0001 from Strem Chemicals composed of 10 nm cobalt nanoparticles in kerosene coated with sodium dioctylsulfosuccinate and a monolayer of LP4 fatty acid condensation polymer. The average volume fraction of cobalt nanoparticles in this ferrofluid is 8.2%. Cobalt has metallic properties (the real part of its refractive index $n$ is smaller than its imaginary part $k$: $n<k$, so that $\varepsilon'=n^2-k^2<0$) in the infrared range, as evident by Fig.1A plotted using data for the optical properties of cobalt reported in [12]. Therefore, self-assembled cobalt nanoparticle chains are suitable for fabrication of hyperbolic metamaterials. When cobalt nanoparticles are completely aligned by the external magnetic field, so that wire array geometry schematically shown in Fig.2A is formed, the diagonal components of the ferrofluid permittivity may be calculated using Maxwell-Garnett approximation as:

$$\varepsilon_z = \alpha\varepsilon_m + \left(1-\alpha\right)\varepsilon_d \qquad (1)$$

$$\varepsilon_{xy} = \frac{2\alpha\varepsilon_m\varepsilon_d + (1-\alpha)\varepsilon_d(\varepsilon_d+\varepsilon_m)}{\left(1-\alpha\right)\left(\varepsilon_d+\varepsilon_m\right)+2\alpha\varepsilon_d} \qquad (2)$$

where $\alpha$ is the average volume fraction of the cobalt nanoparticle phase (assumed to be small), and $\varepsilon_m$ and $\varepsilon_d$ are the dielectric permittivities of cobalt and kerosene, respectively [13]. Wavelength dependencies of $\varepsilon_z$ and $\varepsilon_{xy}$ calculated using eqs.(1,2) at $\alpha=8.2\%$ are plotted in Fig. 1B. While $\varepsilon_{xy}$ stays positive and almost constant, $\varepsilon_z$ changes sign to negative around $\lambda=1\mu m$. Therefore, at $\lambda>1\mu m$ the ferrofluid subjected to external magnetic field becomes a hyperbolic metamaterial, which has been demonstrated in the experiment [1]. Hyperbolic character of the magnetized ferrofluid has been confirmed by polarization-dependent transmission and reflection measurements at single laser lines and broadband FTIR spectroscopy performed in the 500-22000 nm wavelength range [1]. Note that in the near infrared frequency range of interest the magnetic permeability of the ferrofluid may be assumed to be $\mu=1$, and all the off-diagonal terms of the dielectric permittivity tensor of the ferrofluid may be assumed to be zero due to relatively small induced optical activity of kerosene and weak external magnetic field used. Wave equation describing propagation of extraordinary photons inside the ferrofluid may be written in the form of 3D Klein-Gordon equation describing a massive field $\varphi_\omega = E_z$ in 3D Minkowski spacetime:

$$-\frac{\partial^2 \varphi_\omega}{\varepsilon_{xy} \partial z^2} + \frac{1}{(-\varepsilon_z)}\left(\frac{\partial^2 \varphi_\omega}{\partial x^2} + \frac{\partial^2 \varphi_\omega}{\partial y^2}\right) = \frac{\omega_0^2}{c^2}\varphi_\omega = \frac{m^{*2}c^2}{\hbar^2}\varphi_\omega \qquad (3)$$

in which the spatial coordinate $z = \tau$ behaves as a "timelike" variable, and the effective mass equals $m^* = \hbar\omega_0/c^2$. For example, it is easy to check that eq.(3) remains invariant under the effective Lorentz coordinate transformation

$$z' = \frac{1}{\sqrt{1 - \dfrac{\varepsilon_{xy}}{(-\varepsilon_z)}\beta}}\left(z - \beta x\right) \qquad (4)$$

$$x' = \frac{1}{\sqrt{1 - \dfrac{\varepsilon_{xy}}{(-\varepsilon_z)}\beta}}\left(x - \beta\frac{\varepsilon_{xy}}{(-\varepsilon_z)}z\right),$$

where $\beta$ is the effective Lorentz boost. The opposite signs of $\varepsilon_z$ and $\varepsilon_{xy}$ ensure that spatial coordinate z plays the role of time in the Lorentz-like symmetry described by eqs.(4). Instead of the "optical path length" $dl_{opt} = ndl$, which is typically introduced in conventional optics, we may introduce an "optical interval" $ds^2_{opt}$ as

$$ds_{opt}^2 = -\varepsilon_{xy}dz^2 + \left(-\varepsilon_z\right)\left(dx^2 + dy^2\right) , \qquad (5)$$

which remains invariant under the effective Lorentz transformations. Similar to our own 3+1 dimensional Minkowski spacetime, opposite signs of $\varepsilon_z$ and $\varepsilon_{xy}$ in a hyperbolic metamaterial lead to Lorentz symmetry of its effective "optical spacetime". Thus, eq.(3) describes world lines of massive particles which propagate in effective 2+1 dimensional Minkowski spacetime [3,4]. When the ferrofluid develops phase separation into cobalt rich and cobalt poor phases, its microscopic structure and local field intensity in the hyperbolic frequency range may be observed directly using an infrared microscope, so that microscopic properties of the effective Minkowski spacetime defects may be studied in the experiment. Our experiments were performed using a narrow 10 μm pathlength optical cuvette placed under the infrared microscope. The cuvette was filled with the ferrofluid and illuminated from below with a linear polarized 1.55 μm laser. The transmission images of the infrared microscope were studied as a function of magnitude and direction of external magnetic field. The schematic view of our experimental setup is shown in Fig. 1C.

Optical microscope transmission images of the ferrofluid illuminated with 1.55 μm laser before and after application of external magnetic field are shown in Figs. 2BC. While the imaginary part of $\varepsilon_z$ is rather large at $\lambda$=1.55 micron, this was the longest wavelength available to us to perform optical microscopy of the samples. However, the level of metamaterial losses at 1.55 μm was adequate to measure transmission of a 10 micrometer pathlength optical cuvette filled with the ferrofluid. The periodic pattern of self-assembled stripes visible in image C appears due to phase separation. The stripes are oriented along the direction of magnetic field. Quasi three-dimensional representation of this image is shown in Fig.2E. Both images demonstrate an almost defect-free well-ordered periodic alignment of cobalt rich filaments. However, by turning the external magnetic field off and on again multiple times it was possible to create defects in the periodic alignment of these filaments. An example of a microscopic image of such accumulated defects within the volume of a polarized ferrofluid is shown in Fig.2D. While on average the filaments remain aligned along the direction of magnetic field (the same direction as in Fig.2C), multiple defects appear in the filament structure. The complicated topology of the defects is clearly revealed in the quasi three-dimensional representation of this image shown in Fig.2F. It should be pointed out that despite these defects, the measured ferrofluid polarization properties remain the same, indicating that the ferrofluid remains a hyperbolic metamaterial on a large scale. Therefore, these defects play the role of microscopic defects in effective Minkowski spacetime. Let us analyse typical examples of these defects in more detail.

Let us start with some examples of simple point defects ("events" in the effective Minkowski spacetime) shown in Fig.3. Such point defects may arise either due to impurities present in the ferrofluid (as shown in Fig.3A) or due to twisting and/or splitting of the ferrofluid filaments (Fig.3BC). Experimental examples of such defects are indeed easy to find in the microscopic images, as shown in Figs.3D-F. It appears that these defects exhibit some similarity with the proposed point defects of Minkowski spacetime, such as magnetic monopole solutions [6]. Since cobalt nanoparticles possess magnetic dipole moments, which are oriented by the external magnetic field, bending the ferrofluid filaments leads to bending of the magnetic field flux lines, as shown in Figs. 3A-C. As a result, effective "virtual" magnetic charges appear in these locations. Unlike the unperturbed regions of the effective metamaterial 3D Minkowski spacetime, $\nabla_{xy}B$ is nonzero in these locations, where $\nabla_{xy}=(\partial/\partial x,\partial/\partial y)$ is the two-dimensional divergence operator, and z coordinate is identified as a timelike variable. Thus, scattering of extraordinary photons by such point defects may be treated as a model of photon-monopole scattering. Such scattering may be studied in polarization-dependent IR microscopy experiments, as illustrated in Fig.4. However, we must emphasize that a true magnetic monopole would appear as a line (a world line) in 2+1 dimensional spacetime, and the point defects which we observe correspond to virtual monopole-antimonopole pairs.

Measured polarization dependencies of the ferrofluid transmission as a function of magnetic field shown in Fig. 4A are consistent with hyperbolic metamaterial character of ferrofluid anisotropy in the infrared wavelength range at large enough magnetic field. Zero degree polarization corresponds to $E$ field of the electromagnetic wave perpendicular to the direction of external magnetic field (the case of ordinary wave). Such polarization is suitable for imaging the microscopic structure of the ferrofluid. On the other hand, microscopic measurements conducted at 90 degree polarization allow us to assess the effect of microscopic defects on the extraordinary wave propagation. As discussed above (see eq.(3)), propagation of extraordinary photons through the volume of hyperbolic metamaterial may be described in terms of particle world line evolution in effective 2+1 dimensional Minkowski spacetime. Thus, studying extraordinary light scattering by the point defects of the ferrofluid may be treated as a model study of point defects in Minkowski spacetime. An example of such study is presented in Figs. 4BC. The image in Fig. 4B measured at zero degree polarization shows the same point defect (y-split) as in Fig. 3F. The microscopic image of the same region measured at 90 degree polarization clearly demonstrates enhanced extraordinary photon scattering by the ferrofluid filament split. Increased scattering by the effective spacetime defects is indeed expected, since in the absence of such defects the ferrofluid should behave as "empty spacetime".

Let us now proceed with examples of linear and planar defects observed in the bulk of the polarized ferrofluid. It appears that linear defects may behave as particle world lines (see Fig.5A as an example), while planar defects (see Fig.5B) may resemble cosmic strings, which are hypothesized to exist in Minkowski spacetime [5]. While transformation optics-based models of cosmic strings have been proposed [14], their experimental realization was not possible due to difficulties associated with 3D nanofabrication issues. On the other hand, our experimental observations indicate that linear and planar defects in effective Minkowski spacetime appear naturally in the disordered ferrofluid. The cosmic string represents a topological defect of spacetime that may be described in geometrical terms by delta function-valued torsion and curvature components. It has been noticed by several authors [5,15,16] that the spacetime geometries of cosmic strings in 3+1 dimensions exhibit close relations to the distortions of solids, which may likewise be regarded as topological defect lines carrying torsion and curvature. Similar analogies hold in 2+1 dimensional gravity [17], and therefore applicable to the aligned ferrofluid. One of the possible scenarios for such a string analog to appear in the aligned ferrofluid is represented in Fig. 5B. It shows schematically a quasi-3D view of a 1+1 dimensional sheet in space-time, which may form a toy model of a "cosmic string". This structure is based on shifting a twisted linear defect shown in Fig.5A along $y$ coordinate. Such analog of cosmic string may arise due to twisting of the self-assembled cobalt nanoparticle filaments. Observations of the disordered ferrofluid using polarization-dependent IR microscopy presented in Figs. 5C-F indeed reveal multiple linear and planar defects. Comparative examination of these images indicate that twisted filaments observed in zero degree polarization images (Fig. 5E and its zoom in Fig. 5C) give rise to scattering features observed at 90 degree polarization (Fig. 5F). Due to small (10 μm) pathlength of the cuvette, the "side view" images presented in Figs. 5C,E,F do not allow clear differentiation between linear and planar defects of the periodic filament structure. On the other hand, a microscopic image of the ferrofluid sample taken along the axes of the nanowires shown in Fig.5D (which was obtained by directing the external magnetic field perpendicular to the ferrofluid-filled cuvette) enables such differentiation. A planar defect of the ordered filament structure (a grain boundary) is clearly seen in Fig.5D. It is highlighted by the green dashed line, while the broken rows of filaments are indicated by red dashed lines. Thus, planar defects of the ferrofluid filament structure, which correspond to analogs of cosmic strings in effective 2+1 dimensional Minkowski spacetime have been indeed detected in our experiments. Note that the spacetime analogy does help in understanding of our experimental results. Evaluation of Figs. 5E and 5F clearly demonstrates that compared to ordinary photons, the defects of ferrofluid lattice exhibit much stronger effect on extraordinary photon propagation. Unlike ordinary photons, extraordinary photons perceive ferrofluid as an effective Minkowski spacetime. Increased scattering by the defects is indeed expected, since in the absence of such defects from the point of view of extraordinary photons the ferrofluid should behave as "empty spacetime".

Finally, let us examine some volume defects observed in the bulk of the polarized ferrofluid. While several analogs of such defects have been explored in the experiments with plasmonic hyperbolic metamaterials [18,19], reduced dimensionality of such metamaterials severely limits the range of possible configurations. On the other hand, the true 3D nature of newly developed self-assembled ferrofluid-based hyperbolic metamaterials lets us study such interesting new possibilities as the recently proposed spacetime cloak geometry [7]. The spacetime cloak is obtained via $(x,t) \rightarrow (x',t')$ spacetime transformation, which creates a void near the spacetime origin, as illustrated in Fig.6A. The transformation is a composition of a Lorentz boost $(x,t) \rightarrow (\bar{x}, \bar{t})$ with velocity $v = c/n$ (described by eq.(4)), followed by applying a 'curtain map'

$$\bar{x}' = \left[ \frac{\left( \delta + |c\bar{t}| \right)}{\left( \delta + n\sigma \right)} \left( \bar{x} - \text{sgn}(\bar{x})\sigma \right) + \text{sgn}(\bar{x})\sigma \right], \ \bar{t}' \to \bar{t} \ , \qquad (6)$$

followed by an inverse Lorentz transformation $(\bar{x}', \bar{t}') \to (x', t')$ [7]. In the resulting spacetime metric any spacetime event, which occurs inside the spacetime cloak (the black area in Fig.6A) is concealed from a distant observer. Note that the blue arrows in Figs.6AB follow the local $\bar{x}' = const$ direction. In the ferrofluid this direction corresponds to the direction of cobalt nanoparticle filaments. Therefore, macroscopic bending of filaments, which is similar to the bending of blue arrows in Figs. 6AB may lead to appearance of a spacetime cloak geometry in the effective 2+1 dimensional metamaterial Minkowski spacetime. Experimental images of disordered ferrofluid indeed provide multiple examples of such volume defects. An example of such a defect is indicated by the dashed oval in Fig.6C. In order to relate it to eq.(6) a magnified image of the spacetime cloak area indicating its orientation with respect to $(\bar{x}', \bar{t}')$ and $(x', t')$ coordinate planes is shown in Fig.6D. The cloaked area is marked by the green dashed line. The optical field intensity in the defect interior appears to be reduced, which is consistent with the spacetime cloaking effect proposed in [7].

In conclusion, we have presented a systematic microscopic study of point, linear, planar and volume defects in the self-assembled wire array hyperbolic metamaterials based on cobalt nanoparticle filaments aligned by external magnetic field. Since extraordinary light propagation in such metamaterials may be described in terms of effective Minkowski spacetime metric, microscopic defects of the metamaterial exhibit strong similarity with such Minkwoski spacetime defects as magnetic monopoles, cosmic strings and the spacetime cloak. Our experimental observations have many important fundamental implications, which reach far beyond transformation optics and electromagnetic metamaterial theory. As demonstrated in [20,21], nonlinear optics of the polarized ferrofluid may be described in terms of analogue gravity. Moreover, as pointed out recently by Mielczarek [8], the properties of self-assembled magnetic nanoparticle-based hyperbolic metamaterials exhibit strong similarities with the properties of some quantum gravity models, such as loop quantum cosmology. Thus, microscopic studies of polarized ferrofluid provide a unique playground to study a microscopic analogue model of gravity in action.

**Acknowledgments**

This work is supported by the NSF grant DMR-1104676.

**Figure Captions**

**Fig. 1** (A) Real and imaginary parts of $\varepsilon$ for cobalt according to [9]. (B) Corresponding wavelength dependencies of the real parts of $\varepsilon_z$ and $\varepsilon_{xy}$ at $\alpha = 8.2\%$. While $\varepsilon_{xy}$ stays positive and almost constant, $\varepsilon_z$ changes sign to negative around $\lambda = 1\mu m$. (C) Schematic view of our experimental setup.

**Fig. 2.** (A) Geometry of the metal nanowire-based hyperbolic metamaterials. (B-F) Experimentally measured microscopic transmission images of cobalt nanoparticle-based ferrofluid obtained using an infrared microscope and a 1.55 $\mu m$ laser as an illumination source: frames (B) and (C) show microscopic images of the ferrofluid before and after application of external magnetic field. The self-assembled nanowires in (C) are oriented along the direction of magnetic field, so that a hyperbolic metamaterial is formed. Frame (D) shows a microscopic image of the ferrofluid after magnetic field is turned off and on multiple times leading to appearance of defects in the nanowire array structure. While on average the filaments remain aligned along the direction of magnetic field (the same direction as in (C), multiple defects appear in the filament structure. Frames (E) and (F) show quasi-3D representation of images in frames (C) and (D), respectively. While image (E) reveals a well-ordered periodic wire array structure, image (F) shows multiple defects, which appear due to external magnetic field cycling.

**Fig. 3.** (A-C) Examples of the possible point defects in the periodic wire array structure formed by self-assembly of cobalt nanoparticles in external magnetic field. The red arrows represent nanoparticle magnetic moments. The grin circle in (A) represents an impurity. The defect shown in (A) may be treated as a magnetic monopole-antimonopole pair, which resides in the effective 3D Minkowski metamaterial spacetime. The defect shown in (B) represents a simple braid structure, while defect

shown in (C) represents a y-split. Multiple examples of such point defects may be found in Fig.2(F). Magnified experimentally observed images of such point defects from Fig.2(F) are shown in (D), (E) and (F), respectively. The defect areas are marked by circles.

**Fig. 4.** (A) Experimentally measured transmission of the cobalt based ferrofluid at $\lambda$=1.55 μm as a function of external magnetic field and polarization angle. Zero degree polarization corresponds to E field of the electromagnetic wave perpendicular to the direction of external magnetic field. Frames (B) and (C) show microscopic images of a point defect in the ferrofluid measured at 0 and 90 degrees polarization, respectively. Image (C) reveals increased scattering by the point defect.

**Fig. 5.** (A) Schematic geometry of a linear defect in the periodic wire array structure formed by twisting of self-assembled cobalt nanoparticle filaments. Such a linear defect may represent a world line of a particle at rest. (B) Quasi-3D view of a 1+1 dimensional sheet in space-time, which forms one of the possible toy models of a "cosmic string". This structure is based on shifting a twisted linear defect shown in (A) along $y$ coordinate. Only two layers of the 3D structure are shown for clarity. (C) Magnified image of twisted filaments observed in the disordered ferrofluid. (D) A microscopic image of the ferrofluid sample taken along the axes of the nanowires, which was obtained by directing the external magnetic field perpendicular to the cuvette. A planar defect of the ordered filament structure (a grain boundary) is highlighted by the green dashed line. (E,F) Microscopic images of the ferrofluid exhibiting multiple linear and planar defects measured at 0 and 90 degrees polarization, respectively. Image (F) reveals increased photon scattering by these defects.

**Fig. 6.** Example of a volume defect in the effective 3D Minkowski spacetime: (A) Proposed spacetime cloak (or "history editor") geometry based on the curtain transformation [7]. (B) The curtain transformation in the $(\overline{x}', \overline{t}')$ coordinate plane (C) Experimental image of a volume defect (indicated by the dashed oval) in the periodic wire array structure, which closely resembles the spacetime cloak geometry in (A). The optical field intensity in the defect interior is reduced, which is consistent with the proposed spacetime cloaking effect. (D) Magnified image of the spacetime cloak area indicating its orientation with respect to $(\overline{x}', \overline{t}')$ and $(x', t')$ coordinate planes. The cloaked area is marked by the green dashed line.

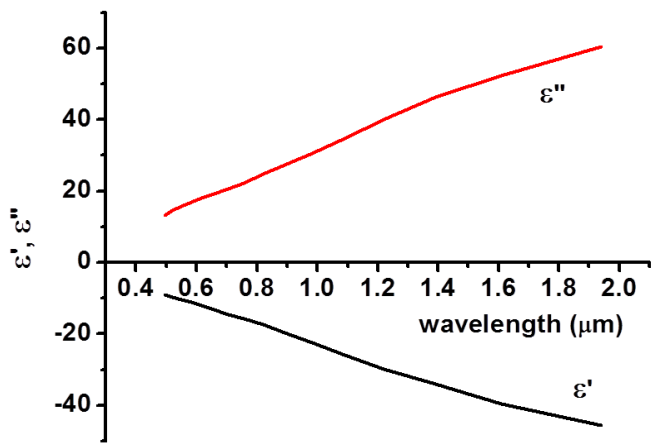

A

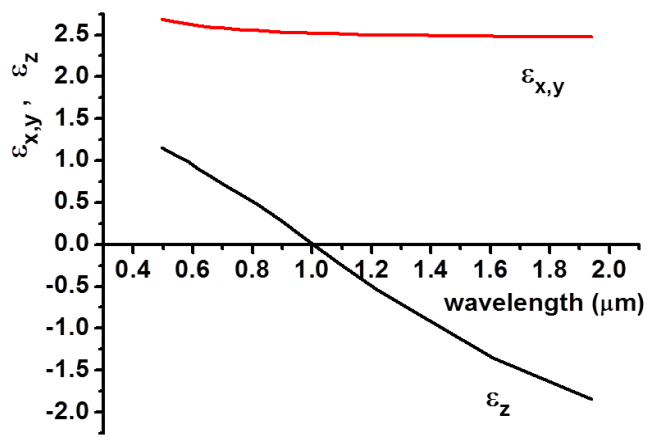

B

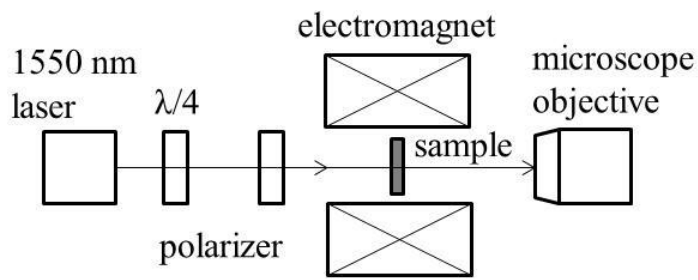

C

**Fig.1**

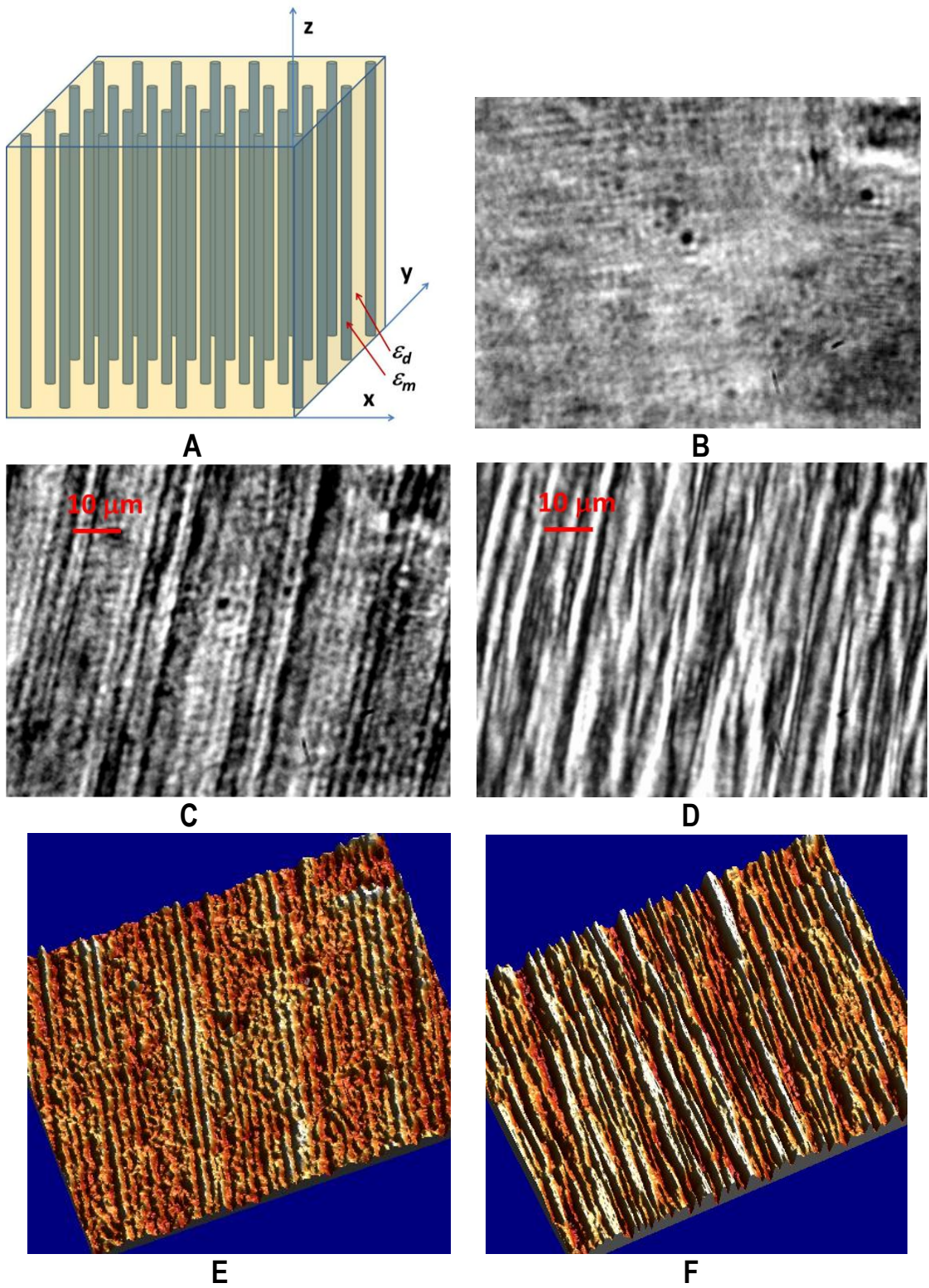

Fig. 2

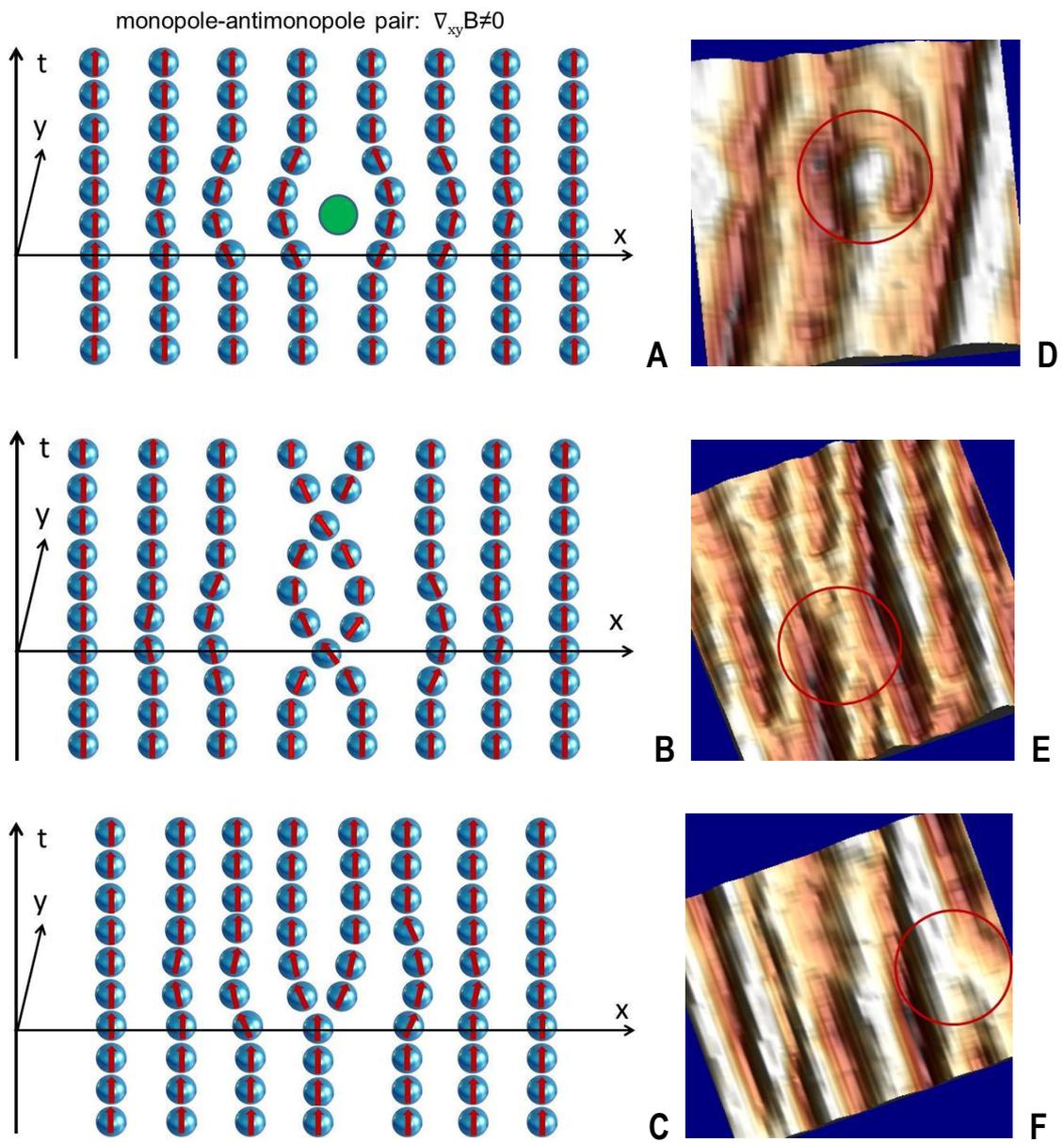

monopole-antimonopole pair: $\nabla_{xy}B \neq 0$

A

B

C

D

E

F

**Fig. 3**

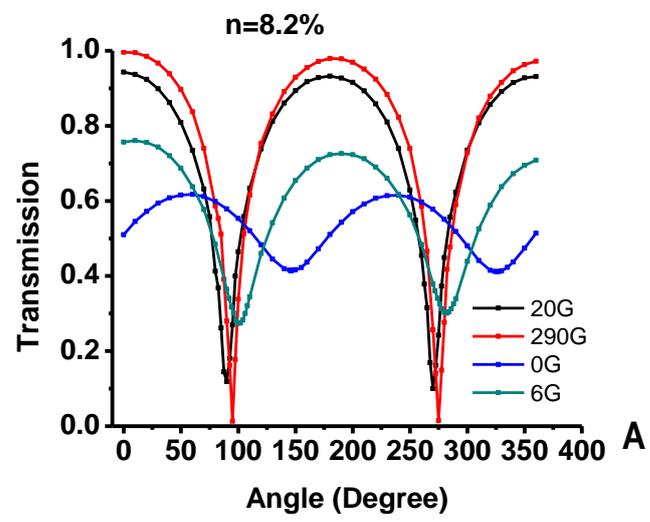

n=8.2%

A

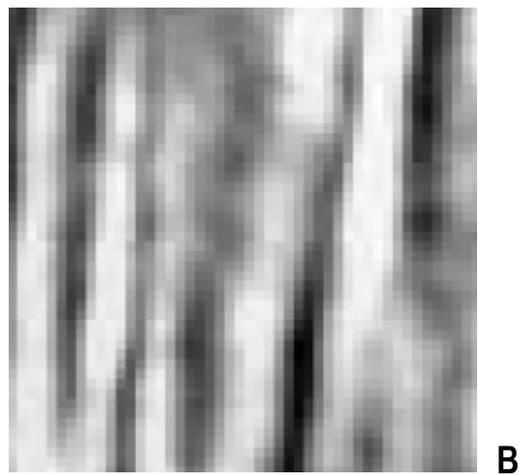

B

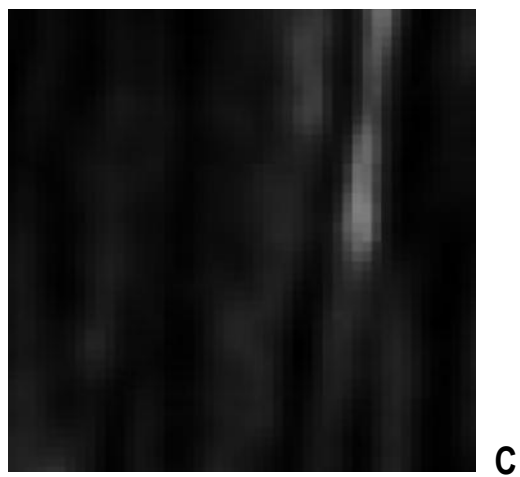

C

Fig.4

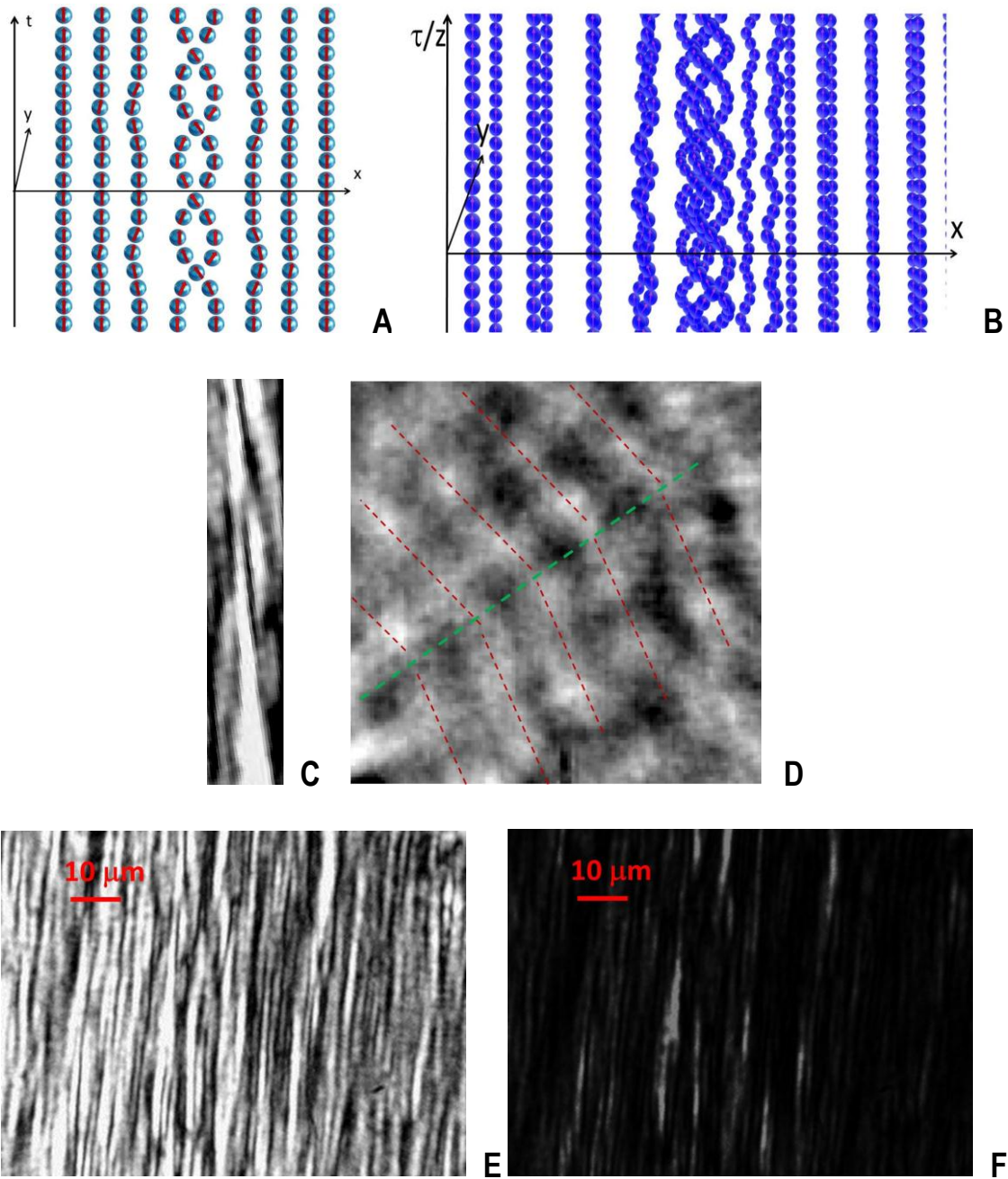

**Fig.5**

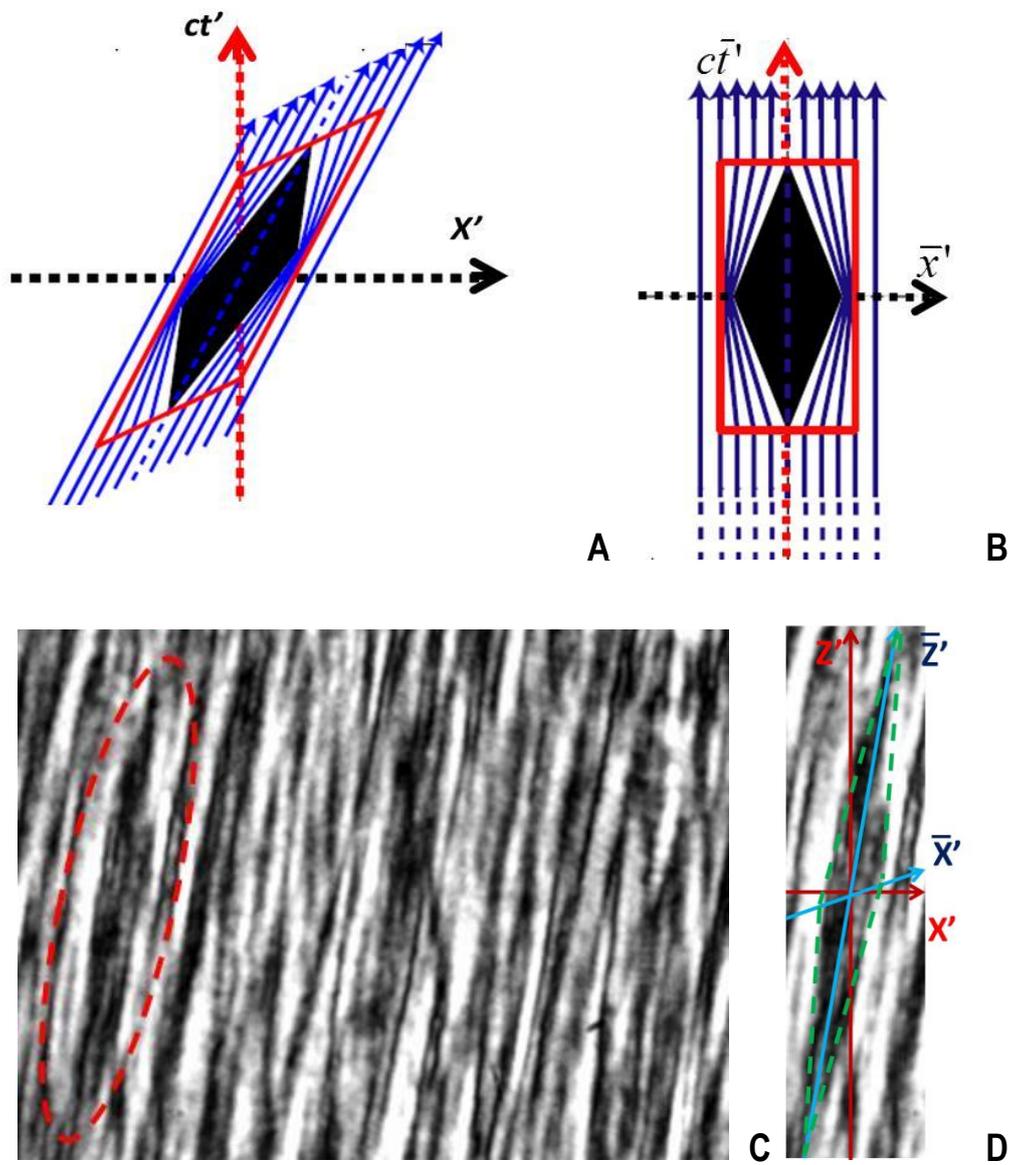

**Fig. 6**